\documentclass[preprint,showpacs,showkeys,amsmath,amssymb,floatfix]{revtex4-1}
\usepackage{graphicx}
\usepackage{bbm}
\usepackage{bm}
\usepackage{color}
\usepackage[pdftex,colorlinks]{hyperref}
\newcommand{\bra}[1]{\ensuremath{\left\langle#1\right|}}
\newcommand{\ket}[1]{\ensuremath{\left|#1\right\rangle}}\hypersetup{urlcolor=blue}
\def\P{\ensuremath{\mathcal{P}}}
\def\id{\ensuremath{\mathbbm{1}}}
\DeclareMathOperator{\Tr}{Tr}
\begin{document}
\title{Statistics of nondemolition weak measurement}
\author{Antonio \surname{Di Lorenzo}}
\email{dilorenzo@infis.ufu.br}
\affiliation{Instituto de F\'{\i}sica, Universidade Federal de Uberl\^{a}ndia,\\
 38400-902 Uberl\^{a}ndia, Minas Gerais, Brazil}
\author{Jos\'{e} Carlos Egues}
\affiliation{Instituto de F\'{i}sica de S\~{a}o Carlos, Universidade de S\~{a}o Paulo}
\begin{abstract}
A measurement consists in coupling a system to a probe and reading the output of the probe  
to gather information about the system. 
The weaker the coupling, the smaller the back-action on the system, but also the less information conveyed. 
If the system undergoes a second measurement, the statistics of the first output 
can be conditioned on the value of the second one. This procedure is known as postselection.
A postselected weak measurement of an observable can give a large average output of the probe 
when the postselected state is nearly orthogonal to the initial state of the system. This large value is an interference effect in the 
 readout of the probe, which is 
initially in a coherent superposition of readout states (also known as pointer states). 
Usually, the weak interaction between system and probe is considered instantaneous, 
so that the dynamics of the probe can be neglected. 
However, for a weak measurement in solid-state devices, an interaction of finite duration is likely needed. 

Here we show how this finite duration generates a contribution of the dynamical phase to the readout statistics. 
Furthermore, we derive interpolation formulas that are able to describe the statistics of the 
weak measurement for the whole range of pre- and postselected states. 
Phase-space averages appear in the expansion, 
suggesting an interpretation in terms of non-positive probabilities. 
Decoherence in the probe is also accounted for and it is pointed out the existence of a regime of intermediate coupling strength 
in which coherent oscillations can be observed in the probability of the readout. 
\end{abstract}
\maketitle
\section{Introduction.}
Sequential or joint nonprojective measurements can reveal the quantum nature of the detectors. For example, 
the pioneering work of Arthurs and Kelly \cite{Arthurs1965} demonstrated how two weak measurements of momentum and position could be carried out simultaneously, and how the uncertainty principle obeyed by the detectors (or probes) showed up in their mutual back-action. While  the theory of positive operator-valued (POV) measures (of which the weak and the strong measurements are particular cases) has its mathematical roots in the seminal papers of Neumark \cite{Neumark1940,Neumark1943} and has been developed since the 70s \cite{Davies1976,Helstrom1976,Holevo1982}, only in 1988  was it realized that a weak measurement followed by a strong one could lead to arbitrarily large values of the average output~\cite{Aharonov1988}, provided that the weak measurement is conditioned on the result of the strong one. This conditioning is called postselection. 
As postselected weak measurements can give an arbitrarily large average output (even for non-orthogonal preparation and 
postselection \cite{DiLorenzo2012d}), they have been used to amplify a weak signal \cite{Hosten2008,Dixon2009,Starling2009}, to settle fundamental issues --- such as determining the traversal time of a barrier \cite{Steinberg1993}, solving Hardy's paradox \cite{Yokota2009,Lundeen2009}, and observing quantum trajectories \cite{Kocsis2011}--- and also to perform quantum state tomography \cite{Lundeen2011}. 
Concerning the last application, there are several proposals \cite{DiLorenzo2011,Haapasalo2011,Lundeen2012,DiLorenzo2012b} for extending the 
procedure to mixed states. It has also been shown \cite{DiLorenzo2004,DiLorenzo2006} that having a stream of particles sent to probes initially prepared to measure in the strong regime can create coherence in the probes; this drives the measurement to the weak regime and is reflected in a deviation of the statistics from that predicted for projective measurement, provided the decoherence rate of the probes does not exceed the firing rate of the particles. 
For other applications and studies of the weak measurement, see Ref.~\cite{Kofman2012}. 

The theoretical papers on postselected weak measurements are mostly limited to the study of the average value 
and assume an instantaneous (von Neumann) interaction. 
While for optical implementations postulating an instantaneous interaction is reasonable, for the still prospective realizations 
in solid-state systems (see the proposals  \cite{Williams2008,Romito2008,Shpitalnik2008,Ashhab2009a,Ashhab2009b,Bednorz2010,Zilberberg2011}) it is more realistic to suppose a coupling that lasts a finite time. 
In an earlier paper \cite{DiLorenzo2008}, we introduced a finite-duration interaction, assuming only that the measurement was a quantum nondemolition one \cite{Braginsky1992}. 
The results of Ref.~\cite{DiLorenzo2008}, however, were limited to the average value of the output and its variance (barring the case study of a spin 1/2, for which, since 
an exact solution exists, the expansion in the weak coupling was made merely to test its validity). In a more recent paper \cite{DiLorenzo2012a}, one of us studied the whole statistics
of the weak measurement, by performing a controlled expansion in the coupling strength, as in Ref.~\cite{Wu2011}, 
and providing an interpolation formula that works for any preparation and postselection. 
Here, we provide the statistics of a nondemolition weak measurement of an arbitrary variable. We show that the finite 
duration has observable consequences, 
since in a weak measurement the coherence of the probe manifests through the contribution of its density matrix off-diagonal elements (in the readout basis), and since the interaction contributes dynamical phases to these elements.  


\section{Description of the measurement.}
Let us consider a quantum system prepared, at time $t_i$, in a 
state ${\rho}_S(t_i)=\rho_i$ (preselection) and a second quantum system, the probe, 
 prepared, at time $t_{0}$, in a state $\rho_P(t_0)=\rho_{0}$. 
Let the Hamiltonian of the system be $\Hat{H}_{S}=\hbar \Hat{\Omega}_S$, and that of the probe $\Hat{H}_P$. 
The system and the probe interact, at time $T_i\ge t_i,t_0$,  
through $H_{int} = -\hbar\lambda g(t) \Hat{x} \Hat{A}$,
where $\Hat{A}$ is an operator on the system's Hilbert space, $\Hat{x}$ on the probe's,  
and $g(t)$ is a function that vanishes outside a finite interval $[T_i,T_f]$, with 
$\int g(t) dt = 1$. The interaction generates a time-evolution operator $\mathcal{U}$ that acts on the Hilbert space of the 
system and the probe, entangling the two. 
Let the operator $\Hat{k}$ be the conjugate observable \footnote{While we are using the notation $x,k$ for the write-in and readout variables, these need not be 
literal position and wave number. Many authors use $p\to x$, $q\to k$, $-\lambda\to \hbar\lambda$.}
of $\Hat{x}$, $[\Hat{x},\Hat{k}]=i$. 
It follows from Heisenberg's equations of motion, or even from Hamilton's equations for the corresponding classical case, 
that $\Hat{k}$ is the observable of the probe that 
carries information about the measured quantity $\Hat{A}$. 
As the interaction lasts a finite time, $\Hat{A}$ has to be be conserved, i.e. $\left[\Hat{A},\Hat{H}_{S}\right]=0$, otherwise the question would arise of what is being measured \footnote{If, instead, we would assume an instantaneous interaction, $T_f\to T_i$, 
$g(t)\to\delta(t-T_i)$, then we could allow $[\Hat{A},\Hat{H}_{S}]\neq 0$.}.  
We notice that, in order for the measurement to be nondemolition, $\Hat{k}$ too must be 
conserved during the free evolution of the probe and change only due 
to the interaction with the observed system. Hence $\Hat{H}_P=\hbar \omega_P(\Hat{k})$.
In other words, we are considering not a von Neumann (instantaneous) weak measurement, 
but the more general nondemolition (finite-duration) weak measurement, having the former as a special limiting 
case. %
At time $t\ge T_f$ a projective measurement 
of the observable $\Hat{k}$ is made on the probe. 
As $\Hat{k}$ is conserved by $\Hat{H}_P$, the value obtained will not 
depend on the time, as long as $t>T_f$. 
In a strong nondemolition measurement, it is furthermore assumed that the probe is prepared in a state 
such that  the elements 
$\bra{k,a}\mathcal{U} \rho_0 \mathcal{U}^\dagger\ket{k,a}$ are negligible unless $k\simeq k(a)$, where the function $k(a)$ maps the values of $a$ to the readout. This hypothesis guarantees 
the fidelity of the measurement, as well as its repeatability. However, as we are interested in the weak regime, we drop this assumption. 

\section{Postselection in a mixed state.}  
\begin{figure}
\includegraphics[height=3in,width=7in]{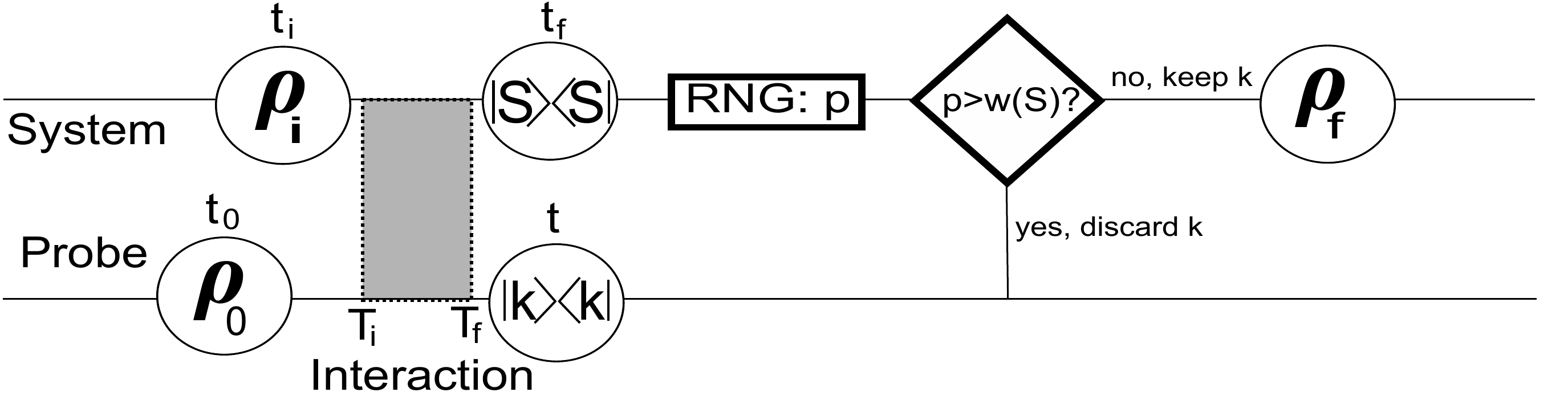}
\caption{\label{fig:scheme}
A schematic view of the measurement with pre- and postselection, the horizontal direction 
representing increasing time. A random number generator yields $0\le p\le 1$ after each trial. If $p\le w(S)$ the outcome 
is considered. 
}
\end{figure}

At time $t_f\ge T_f$, a projective measurement of an observable $\Hat{S}$ of the system is made, 
giving an output $S$ and leaving the system in the state $|S\rangle$. 
Then, given $S$, one keeps the outcome $k$ according to an arbitrarily 
chosen probability $w(S)$. 
This step leaves the system in the postselected mixed state 
\begin{equation}
\rho_f=\frac{\sum_{S} w(S) \ket{S}\bra{S}}{\sum_S w(S)}.
\end{equation} 
The procedure detailed above and sketched in Fig.~\ref{fig:scheme}, 
describes a measurement with pre- and postselection. 
Without loss 
of generality, we shall consider $t_i=t_0=T_i=0$ and $t_f=T_f=\tau$, with $\tau$ the duration of the interaction. 
We notice that $W\equiv \sum_S w(S)$ is not necessarily 1 as $w(S)$ are probabilities \emph{conditional} on the event $S$, not probabilities \emph{of} the event. 
If the system has a finite-dimension Hilbert space, with $D$ the dimension, 
then $0\le W\le D$. 
For 
$
w(S)=
w_0$ if $S=S_0$ and 0 otherwise, 
the postselection is in the pure state $\rho_f=\ket{S_0}\bra{S_0}$. 
In this case, $w_0<1$ is a sub-optimal choice, in the sense that some of the trials 
are discarded unnecessarily. The opposite limit is found for 
$w(S)=w=W/D
 \, , \forall S$.
This implies that no postselection is made, i.e., $\rho_f=D^{-1}\id$. The optimal choice is $w=1$. 
If all the probabilities are multiplied by the same factor $r$, $r w(S)\to w(S)$, the postselected state is unchanged, 
but the probability of a successful postselection is also multiplied by $r$, and so is the joint probability of observing  the probe in $p$ and successfully post-selecting the system. 
Thus the conditional probability, given by the ratio of the two probabilities above, is unaffected by this rescaling. 
The optimal choice of $r$ is such that it maximizes the 
probability of postselection, namely  $\max_S\left[{w(S)}\right]=1$. Hence $W_{opt}\ge 1$, with the equality only for postselection in a pure state. 
In a sense, we could say that the probabilistic postselection leaves the system in the unnormalized state 
$\sum w(S) \ket{S}\bra{S}$.

Another proposed method \cite{Wiseman2002} is to make the postselecting measurement a POV one. 
This way $\rho_f$ is replaced by a positive operator $\Hat{E}_f$. 
However, while $\Hat{E}_f$ appears in the probabilities, as e.g. in Eq. \eqref{eq:instgenprob} below, 
the system is not selected in a state $\Hat{E}_f/\mathrm{Tr}\Hat{E}_f$ but in $\rho_f =\Hat{E}_f^{1/2} \rho'_i  \Hat{E}_f^{1/2}$, with $\rho'_i$ its reduced density matrix after the interaction with the probe. 
%

\section{Exact results.}
The joint state for the probe and the system, at any time $t\ge \tau$ is
\begin{align}
\nonumber
\rho(a,a',k,k';t) &=\! \int dk_1 dk_2 \ 
\rho_{0}(k_1,k_2) {\rho}_{i}(a,a')\bra{k,a} \mathcal{U}_{0,\tau} \ket{k_1,a}
\bra{k_2,a'} \mathcal{U}_{0,\tau}^\dagger\ket{k',a'} 
\\
&\times\exp\biggl(\!i
\bigl\{[\omega_P(k')-\omega_P(k)]+[\omega_S(a')-\omega_S(a)]\bigr\}(t-\tau)\biggr)
,
\label{eq:finstate}
\end{align}
where $\mathcal{U}$ is the time evolution operator generated by $H_{S}+H_P+H_{int}$, 
$|k,a\rangle$ are the simultaneous eigenstates of $\Hat{k}$ and $\Hat{A}$,  
and $\omega_S(a)$ are the eigenvalues of $\Hat{\Omega}_S$ corresponding to $|a\rangle$. 
Following a lemma demonstrated in the Supplemental Material section, there is an analytic solution for the propagator, 
\begin{align}
&\bra{k,a} \mathcal{U}_{0,\tau} \ket{k_0,a_0}
=\delta_{a,a_0}\delta\left(k-k_0-\lambda a\right)e^{-i\Gamma_a(k)} ,
\label{eq:prop}
\end{align}
where we define  
the Hamiltonian phase 
\begin{equation}\label{eq:genhamphase}
\Gamma_a(k):=\int_{0}^{\tau}\!\!\!ds\ \omega_P\!\left(k-\lambda\; a\left[1-\!h(s)\right]\right),
\end{equation}
with $h(s)=\int_0^s ds' g(s')$. 
The joint probability of observing the outcome $k$ at time $t$ for the probe and 
of postselecting the system in $\rho_f$
at time $T_f$, follows readily from Eqs.~\eqref{eq:finstate} and \eqref{eq:prop}:  
\begin{align}
\P(k,{\rho}_f) =&  
W\sum_{a,a'}
\bra{a'}\rho_f \ket{a} \bra{a}\rho_i\ket{ a'} e^{-i\left[\Gamma_{a}(k)-\Gamma_{a'}(k)\right]} 
\label{eq:instgenprob}
\rho_0(k-\lambda a,k-\lambda a')
.
\end{align}
For a von Neumann measurement, $\Gamma_a(k)=0$. 

The conditional probability of observing the outcome $k$, given that the state 
has been postselected in ${\rho}_f$, follows from Bayes's rule, 
\begin{align}
\label{eq:genweakprob}
\mathcal{Q}(k):= 
 \frac{\P(k,{\rho}_f)}{\P_{post}},
\end{align}
where the denominator $\P_{post}=\int dk\, \P(k,{\rho}_f )$ 
represents the probability of making a successful postselection in 
${\rho}_f$, irrespective of the value of $k$, or of what observable of the probe, if any, was measured. 
What remains to be done is simple: apply the controlled expansion of both 
$\mathcal{P}(k,\rho_f)$ and $\mathcal{P}_{post}$ in $\lambda$ 
as in Ref.~\cite{DiLorenzo2012a}, with the difference that here one should keep track of additional contributions from the dynamical phases $\Gamma$.
It is fundamental to keep in mind that, both in $\P_{post}$ and in $\P(k,\rho_f)$, the zeroth and first order terms vanish for  
nearly orthogonal pre- and postselected states (NOPPS). Thus one should write down $\mathcal{Q}(k)$ as the ratio of two quadratic polynomials in $\lambda$, 
without succumbing to the temptation to expand the denominator. 
%
\section{Definitions.}
Before proceeding to the expansion, 
we introduce the normal weak values as in Ref.~\cite{DiLorenzo2012a}
\begin{align}
&\alpha_{m,n}\equiv\mathrm{Tr}_{S}\{\Hat{A}^m\rho_f \Hat{A}^n \rho_i \} . 
\end{align}
We note that $\alpha_{n,m}=\alpha_{m,n}^{*}$, so that $\alpha_{m,m}$ are real. 
For NOPPS, $\alpha_{0,0}\to 0$ and $\alpha_{0,n}\to 0$, with $\alpha_{0,0}/\alpha_{0,n}\to 0$ for $n\ge 1$; also,  
$\alpha_{m,n}\to \mu\neq 0$ for $m,n\ge 1$.  While there can be exceptions to this behavior when  $\Hat{A}^{n+1}=\Hat{A}$ for some integer $n$, 
in any case $\alpha_{1,1}$ stays finite for NOPPS, barring some trivial instances.  
This consideration is important, as $\alpha_{1,1}$ provides the dominant term to both $\P(k,\rho_f)$ and $\P_{post}$ 
for NOPPS. 

We shall also use the phase-space averages \cite{Moyal1949}, 
$\overline{f(x,k)}\equiv \int dk dx\, f(x,k) \Pi^W_0(x,k)$, with 
$\Pi^W_0$ the initial Wigner function of the probe. 
In particular, we define the covariance $C(f,g)=\overline{f g}-\overline{f}\, \overline{g}$. 
A natural concept arising in the expansion is that of phase-space conditional averages $\overline{f(x,k)}_{|k}$, defined by  
\begin{equation}
\overline{f(x,k)}_{|k} \P_0(k) =\int dx\, f(x,k) \Pi^W_0(x,k)
\label{eq:condphsp}
\end{equation}
with $\P_0(k)=\rho_0(k,k)$. 
The phase-space average of $f(x,k)$ is found by integrating Eq.~\eqref{eq:condphsp} over $k$. 
Finally, we define the time scales 
$\tau_n=\int_0^\tau ds\, [h(s)]^n [1-h(s)]$, which satisfy $\tau_n\ge \tau_{n+1},  n\in \mathbb{N}$. 

%

\section{Weak compared to what?}
As $\lambda$ is a dimensionful constant, it has to be compared 
to some homogeneous quantity in order to establish whether the measurement is weak, strong, or intermediate.  
Let  $\kappa_k$ be the coherence scale, i.e., $k\gg \kappa_k\implies \rho_0(K+k,K-k)\ll \rho_0(K,K)$, 
and $\Delta_k$ the classical uncertainty scale, i.e., $k\gg \Delta_k\implies \rho_0(k,k)\ll \rho_0(0,0)$. 
Precisely, we define $\kappa_k=(\Delta_x)^{-1}$, where $\Delta_x$ is the classical uncertainty of the conjugate variable of $k$, $\Delta_x^2 = \overline{x^2}-\overline{x}^2$, 
and 
$\Delta_k^2 = \overline{k^2}-\overline{k}^2$. 
The uncertainty relation requires $\kappa_k\le 2\Delta_k$. 
In order to realize the weak (coherent) regime, the coupling constant must be small compared to the coherence scale, precisely,  $\lambda a_M\ll \kappa_k$, with $a_M$ the maximum distance between the eigenvalues of $\Hat{A}$. The validity of the expansion relies also \cite{Wu2011} on $\lambda \overline{x}\ll 1$. 
Equation~\eqref{eq:instgenprob} shows that both the off-diagonal terms of the detector 
and the Hamiltonian phase contribute to the statistics. 
Thus we are in presence of interference. 
We also remark that if either $\rho_f\propto\id$ or $\rho_i\propto\id$, i.e., no postselection or no preselection is made, the coherent contributions disappear. 
Hence, both pre- and postselection are essential in order for interference to show up. 
%

\section{Controlled expansion.}
The probability of postselection is found by expanding the integral of Eq.~\eqref{eq:instgenprob} 
(for simplicity, $W=1$), 
\begin{align}
&P_{post}\simeq\ \alpha_{0,0} - 2\lambda\, \overline{x\vphantom{l}}_{\tau_0}\, \mathrm{Im}(\alpha_{0,1})+\lambda^2\overline{x_{\tau_0}^2}\alpha_{1,1} .
\end{align}
where $x_t=x+\omega_P'(k) t$ and the prime stands for differentiation. 
Notice that $\omega_P'(k)$ is the velocity, so that $x_t$ is the displacement of the $x$ variable. 
Due to its non-trivial dynamics as $k$, and hence the velocity, changes during the interaction with the system, 
$\tau_0$ appears instead of $\tau$ \footnote{For simplicity, consider a constant interaction, so that $k$ varies linearly in time for fixed $a$ (uniformly accelerated motion). Then $\tau_0=\tau/2$, and the classical displacement 
is indeed $x_\tau=x_0 + v_\tau \tau/2$. Notice that the $k$ appearing in the formulas refers to the value at time $\tau$, hence 
$\omega'(k)$ is the final velocity $v_\tau$.}.
Since the second-order term contributes significantly only for NOPPS, we have neglected $\alpha_{0,2}$ compared to $\alpha_{1,1}$. 

Furthermore, as all expressions are homogeneous in $\alpha_{m,n}$, we may reduce the number of independent parameters 
by dividing by, say, $\alpha_{0,0}$. Accordingly, we introduce the canonical complex weak value  $A_w=\alpha_{0,1}/\alpha_{0,0}$ 
and the additional real term $B_w=\alpha_{1,1}/\alpha_{0,0}$. We remark that $B_w\ge |A_w|^2$. 
The equality holds whenever $\rho_f$ and $\rho_i$ are mixtures of pure states $\sum_fw_f\ket{f}\bra{f}$ and  
$\sum_iw_i\ket{i}\bra{i}$ each pair of which has the same weak value $\bra{f}\Hat{A}\ket{i}/\bra{f}\ket{i}=A_w\ \forall i,f: w_i w_f\neq 0$. 
In particular, $B_w=|A_w|^2$ for pure pre- and postselected states. 

It is convenient to introduce the characteristic function $Z(\theta)\!=\!\int dk\, e^{i\theta k} \mathcal{Q}(k)$, from which 
the moments of the distribution can be generated as $\langle k^n\rangle\!=\!\left(-i\partial / \partial\theta\right)^n Z(\theta)|_{\theta=0}$, 
with $\langle\cdots \rangle$ average over $\mathcal{Q}$. 
By integrating Eq.~\eqref{eq:instgenprob} times $\exp[i\theta k]$, changing the variable of integration to $k-\lambda(a+a')/2$, expanding, and normalizing,  
\begin{align}
NZ(\theta)\simeq\ &\overline{e^{i\theta k}}
+\lambda 
\left[i\theta\overline{e^{i\theta k}}\mathrm{Re}(A_w)-2\overline{e^{i\theta k\vphantom{^2}}x_{\tau_0}} 
\mathrm{Im}(A_w)\right]
+\lambda^2 \overline{e^{i\theta k}\left(x_{\tau_0}^2+\frac{\theta^2}{4}\right)}
B_w ,
\label{eq:charfunc}
\end{align}
with 
\begin{align}
\label{eq:postinterpol}
&N=\frac{P_{post}}{\alpha_{0,0}}\simeq\ 1- 2\lambda\, \overline{x_{\tau_0}\vphantom{l}}\, \mathrm{Im}(A_w)+\lambda^2 \overline{x_{\tau_0}^2}B_w .
\end{align}
The interpolating formula for the probability is instead 
\begin{align}
N\mathcal{Q}&= \P_0\left[1-2\lambda \overline{x}_{\tau_{0}|k}\mathrm{Im}(A_w)
+\lambda^2
\overline{x^2_{\tau_0}}_{|k} B_w\right]
-\lambda \left[ \P_0'\,\mathrm{Re}(A_w)-\frac{\lambda}{4}\P_0''B_w
\right] .
\label{eq:naive}
\end{align}
The derivatives come from having expanded $\rho_0(k-\lambda a,k-\lambda a')$. This way, the shifts of the probability distribution are lost. 
However, as in the weak regime the shifted peaks are not resolved, this yields a small deviation from the exact result.

As an example, away from NOPPS, we can expand the average value of $k$ to first order, 
\begin{equation}
\langle k\rangle\simeq \overline{k} +\lambda\mathrm{Re}(A_w)-2\lambda C(x_{\tau_0},k) \mathrm{Im}(A_w).
\end{equation}
The contribution from $\mathrm{Im}(A_w)$ is usually neglected, as in an instantaneous measurement $x_{\tau_0}=x$ and 
as a rule $x$ and $k$ are assumed to be initially uncorrelated. The correlator $C(x, k)$ 
is proportional to the derivative of the variance of $k$ under specific hypotheses \cite{Jozsa2007}.

Furthermore, it may be useful to observe the value $x$ of the probe, instead of $k$. We give the full statistics in the Supplemental Materials section, and give here the average value of $x$, which is, in the regime away from NOPPS, where we are allowed to keep up to first order terms, 
\begin{equation}
\langle x\rangle \simeq \overline{x}_\tau-2\lambda \left[C(x_\tau,x_{\tau_0})
\mathrm{Im}(A_w)
-\overline{\omega_P''}t_v \mathrm{Re}(A_w)\right].
\end{equation}
The results of Ref. \cite{Jozsa2007} are recovered as a special case.
\begin{figure}[t!]
\centering{
\includegraphics[width=2.7in]{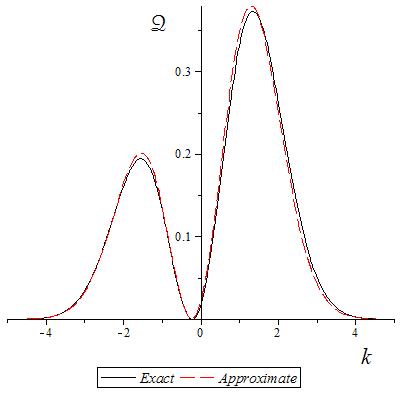}}
\caption{
\label{fig:prob}
Comparison between the interpolating (dashed) and the exact formula (solid) for the probability $\mathcal{Q}(k)$ 
of obtaining $k$ in a weak measurement of $\Hat{A}=\Hat{\sigma}_z$.  The preselection is in a pure state oriented at an angle $\pi/3$ with the 
$z$ axis, and the postselection makes an angle $\pi-0.1$ with the preselection and lies in the plane defined by $\mathbf{m}$ and  $\mathbf{a}$. 
All variables are in units of $\Delta_k$: $\lambda=0.5$, $\kappa_k=2$, $k_H=10$, $k_D=\infty$. 
A not so weak coupling strength was chosen, in order to have a discernible difference between the two curves.}
\end{figure}

%
\begin{figure}[t!]
\centering{
\includegraphics[width=2.7in]{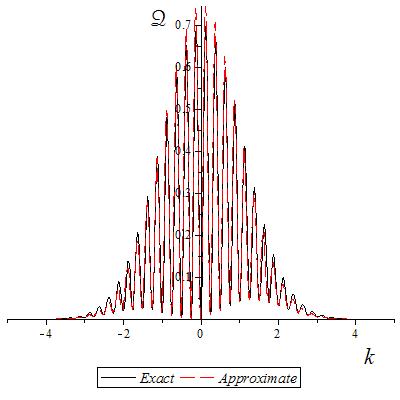}}
\caption{
\label{fig:osc}
Coherent oscillations of the readout probability. Here, 
$k_H=0.2$ and the other parameters are as in the previous figure.}
\end{figure}
\section{Decoherence.}
As an effective model for decoherence, we consider a random classical force acting on the probe, while the latter interacts with the system. 
In principle, if the probe is prepared before the interaction starts, decoherence will tend to kill the off-diagonal elements $\rho_0(k,k')$, driving the 
measurement to the weak incoherent regime. However, this effect can be counteracted by preparing the probe shortly before its interaction with the system, and 
in any case it can be treated rather simply. On the other hand, decoherence during the interaction with the system is unavoidable, as the probe is open 
to external influences, which can come from the system and from the environment. 
For simplicity, we take $\omega_P(k)=\hbar k^2/2M$. The effect consists in the addition of an imaginary phase to $\Gamma_a-\Gamma_{a'}$ in Eq.~\eqref{eq:instgenprob}, 
which becomes
\begin{align}
 \P(k,{\rho}_f) =&  
\sum_{a,a'}
\bra{a'}\rho_f \ket{a} \bra{a}\rho_i\ket{ a'} e^{-i\left[\Gamma_{a}(k)-\Gamma_{a'}(k)\right]} 
\label{eq:instgenprob2}
 e^{-\lambda^2 (a-a')^2 \gamma k_B T \varepsilon \tau^3/M}\rho_0(k\!-\!\lambda a,k\!-\!\lambda a')
.
\end{align}
We assumed the random force to have the correlator $\langle f(t)f(t')\rangle=\delta(t-t')  \mu k_B T$, with $\mu$ a constant, $T$ the temperature, and $k_B$ Boltzmann's constant, 
and we defined the decoherence rate $\gamma=\mu/M$ \footnote{Strictly speaking, as $k$ is not necessarily a wave number, $M$ does not have the dimensions a mass, and hence 
$\gamma$ does not have the dimensions of a rate.}. 					
The factor $\varepsilon$ is a dimensionless parameter that depends on the details of $g(t)$. Thus there is a new scale for $\lambda$ to be compared with, 
the decoherence scale $K_D=[\gamma k_B T \tau^3/2M]^{-1/2}$. 
If $\lambda\gg K_D$, the contribution from the off-diagonal elements of $\rho_0(k,k')$ becomes negligible, and we fall into the weak incoherent regime. 
If $\lambda\ll K_D$, the net result of decoherence is simply to shift the coefficients of $B_w$ in Eqs. \eqref{eq:charfunc} and \eqref{eq:naive}.
%

\section{A case study.} 
We consider a spin-1/2 system preselected in a state $\rho_i=(1+\mathbf{m}\cdot\boldsymbol{\sigma})/2$, 
 on which a weak measurement of $\Hat{A}=\mathbf{a}\cdot\boldsymbol{\sigma}$ is made, and that is 
postselected in $\rho_f=(1+\mathbf{n}\cdot\boldsymbol{\sigma})/2$. 
The interaction is considered constant $g(t)=\tau^{-1}$ and the probe is prepared in 
$\rho_0(k,k')\propto \exp{\{-(k+k')^2/8\Delta_k^2-(k-k')^2/2\kappa_k^2\}}$. 
The free Hamiltonian of the probe is as in the previous section $\hbar^2 k^2/2M$. This defines the Hamiltonian 
scale $k_H = \sqrt{2M/\hbar\tau}$. 
We show the approximate and exact probability in Fig.~\ref{fig:prob}. 

In the intermediate case $k_H^2/\Delta_k\ll\lambda\ll \kappa_k$, it is no longer legitimate to expand $\exp{(i\Gamma_a)}$. Then, 
as shown in Fig.~\ref{fig:osc}, the probability displays coherent oscillations in $k$. 
%

\section{Conclusions.} We have studied the statistics of a weak nondemolition measurement, providing expressions for the 
probability and the characteristic functions that are robust for any overlap between the postselection and the preparation, 
contrary to the results of Ref.~\cite{Aharonov1988} and subsequent papers on weak measurement. We have included decoherence in an effective, albeit heuristic, way, and we have pointed out the existence of a regime of intermediate strength, in which coherent oscillations can be observed. 

\begin{acknowledgments}
A.D.L. performed this work as part of the Brazilian Instituto Nacional de Ci\^{e}ncia e
Tecnologia para a Informa\c{c}\~{a}o Qu\^{a}ntica (INCT--IQ) and 
was supported by Funda\c{c}\~{a}o de Amparo \`{a} Pesquisa do 
Estado de Minas Gerais through Process No. APQ-02804-10.
\end{acknowledgments}
%

\appendix
\section{A useful lemma}
We prove the following lemma: 
for a system subject to the time-dependent Hamiltonian 
$\Hat{H}=\hbar\omega(\Hat{k})-f(s) \Hat{x}$, with $\omega,f$ arbitrary functions and 
$[\Hat{x},\Hat{k}]=i$, with the time-evolution being hence
$\mathcal{U}=\mathcal{T} \exp{\left\{-i\int_0^t d s \left[\omega(\hat{k})-f(s) \hat{x}\right]\right\}}$, 
the propagator reads  
\begin{align}
&\langle k | \mathcal{U}|k_0\rangle 
=\delta\!\left(k\!-\!k_0\!-\!\int_0^t ds f(s)\right)
\exp{\left[-i\int_0^t ds\, \omega\!\left(k-\int_s^{t} ds' f(s')\right)\right]}.
\end{align}
\subsection{Brute force derivation}
We use the path-integral technique, but without path-integrals, i.e. we approximate the time-ordered exponential 
as a product of $N+1$ terms, and we introduce between each term the identity in the $|k\rangle$ basis, 
obtaining ($k_{N+1}\equiv k$)
\begin{align}
\nonumber
\langle k | \mathcal{U} |k_0\rangle 
&\simeq	 \int dk_1\cdots dk_N\prod_{j=0}^{N}  \langle k_{j+1} | \exp{\left\{-i\varepsilon_j  \left[\omega(\Hat{k})-f(s) \Hat{x}\right]\right\}}  |k_j\rangle 
\\
\nonumber
&\simeq  \int dk_1\cdots dk_N\prod_{j=0}^{N}  \langle k_{j+1} | 
\exp{[-i\varepsilon_j  \omega(\Hat{k})]}
\exp{[i\varepsilon_jf(t_j) \Hat{x}]}  |k_j\rangle 
\\
\nonumber
&= \int dk_1\cdots dk_N\prod_{j=0}^{N} 
\delta\!\left(k_{j+1}-k_j-\varepsilon_jf(t_j)\right) 
\exp{[-i\varepsilon_j  \omega(k_{j+1})]}\\
\label{eq:pathint}
&=\delta\left(k-k_0-\sum_{j=0}^{N}\varepsilon_j f(t_j)\right) 
\exp{\left\{-i\left[\sum_{j=0}^{N}\varepsilon_j\ \omega\!\left(k_0+\sum_{m=0}^{j}\varepsilon_m f(t_m)\right)\right]\right\}}.
\end{align}
In the limit $N\to\infty$, the sums in the last line of Eq. \eqref{eq:pathint} become integrals, and the 
approximated equality with the first line becomes exact, thus the lemma is proved.
\subsection{Alternative derivation}
We provide an alternative, more elegant derivation of the above result. The technique illustrated below may find 
applications in other fields. 
The Schr\"{o}dinger equation in wave number space is 
\begin{equation}\label{eq:schreq}
\omega(k) \psi(k,t)+if(t) \frac{\partial}{\partial k}\psi(k,t)=i\frac{\partial}{\partial t} \psi(k,t) .
\end{equation}
Upon rearranging the terms and dividing by $\psi(k,t)$, we have the non-homogeneous first order partial differential equation
\begin{equation}
\left[f(t) \frac{\partial}{\partial k}-\frac{\partial}{\partial t}\right] u(k,t) =i\omega(k) ,
\end{equation}
with $u=\ln{\psi}$. The solution to the corresponding homogeneous equation is 
$u_0\left(k+\int_0^t f(s) ds\right)$, with $u_0$ an arbitrary function of one variable.\footnote{We may change the lower limit of integration to an arbitrary constant $c$, but this 
gives an additional constant $C=\int_0^c f(s) ds$ in the argument of $u_0$ that can be reabsorbed in the arbitrary function $u_0$. Thus we chose zero as lower limit, just because we like it.} 
This suggests to change the variables to $\mu,t$, with  $\mu=k+\int_0^t  f(s)ds$. Let $v(\mu,t)=u(k(\mu,t),t)$. The PDE becomes then
\begin{equation}
-\frac{\partial}{\partial t} v(\mu,t) =i\,\omega\!\left(\mu-\int_0^t f(s)ds\right) ,
\end{equation}
and a particular solution is readily found 
\begin{equation}
 v(\mu,t) =-i\int_0^t \omega\!\left(\mu-\int_0^s f(s')ds'\right)   ds,
\end{equation}
so that, in terms of the original function, the general solution is 
\begin{equation}
u(k,t) =-i\int_0^t \omega\!\left(k+\int_s^t f(s')ds'\right)   ds + u_0\left(k+\int_0^t f(s) ds\right).
\end{equation}
The arbitrariness of $u_0$ can be exploited to find the solution of Eq.~\eqref{eq:schreq} with the initial 
condition $\psi(k,0)=\psi_0(k)$: 
\begin{equation}
\psi(k,t) =\exp{\left\{-i\int_0^t \omega\!\left(k+\int_s^t f(s')ds'\right)   ds\right\}}\, \psi_0\left(k+\int_0^t f(s) ds\right).
\end{equation}

\section{Statistics of the write-in variable}
The characteristic function $\check{Z}(\chi)\equiv \int dx \exp{(i\chi x)} \check{\P}(x|\rho_f)$ is 
\begin{align}
\nonumber
\check{Z}(\chi)\propto&\ \overline{\exp{\left(iF_{\chi,\tau}\right)}}-2\lambda\left[\overline{x_{\chi,\tau_0} 
\exp{\left(iF_{\chi,\tau}\right)}} \mathrm{Im}(A_w)
-i \overline{y_{\chi,t_v} \exp{\left(iF_{\chi,\tau}\right)}}\mathrm{Re}(A_w)\right]
\nonumber
\\
&
+\lambda^2 \left[\overline{\left(x_{\chi,\tau_0}^2-y_{\chi,t_v}^2\right) \exp{\left(iF_{\chi,\tau}\right)}}
-\frac{i}{2} \overline{\left(\partial_k^2 F_{\chi,\tau_0} \right)\exp{\left(iF_{\chi,\tau}\right)}}\right] B_w,
\label{eq:chix}
\end{align}
with  $F_{\chi,t}=\chi x+[\omega_P(k+\chi/2)-\omega_P(k-\chi/2)]t$, 
$x_{\chi,t}=\partial_\chi F_{\chi,t}$, 
$y_{\chi,t}=\partial_k F_{\chi,t}$, 
$t_v=(\tau-\tau_0)/2$. The normalization is given by  $N=\P_{post}/\alpha_{0,0}$.
We note that $\overline{\exp{\left(iF_{\chi,\tau}\right)}}$ is the characteristic function for $x$ that the probe would have after 
a time $\tau$, had it not interacted with the system. 

\section{Measurement of a spin 1/2}
\subsection{Exact expressions}
We use units of $\Delta_k$, $k/\Delta_k\to k$, $\lambda/\Delta_k\to \lambda$, $\theta \Delta_k\to \theta$. 
We assume the free Hamiltonian for the probe to be $\Hat{H}_P=\hbar^2\Hat{k}^2/2M$. 
This position defines the dynamical scale: $k_H^2=M/\hbar \tau_0$, with $\tau_0=\int_0^\tau dt \int_t^\tau ds\, g(s)$. 
\begin{align}
\P(k,\rho_f)&=\sum_\sigma \biggl[
\rho_0(k-\lambda\sigma,k-\lambda\sigma) \bra{\sigma}\rho_f\ket{\sigma} \bra{\sigma}\rho_i\ket{\sigma}
\nonumber
\\
&+
e^{2i\sigma \lambda k/k_H^2}\rho_0(k-\lambda\sigma,k+\lambda\sigma) \bra{-\sigma}\rho_f\ket{\sigma} \bra{\sigma}\rho_i\ket{-\sigma}
\biggr]
\end{align}
For definiteness $\rho_0(k,k')\propto \exp{\{-[(k+k')^2/8-(k-k')^2/2\kappa_k^2]\}}$, with $\kappa_k\le 2$ the coherence scale. 
We let $\rho_i=(1/2) (1+\mathbf{m}\cdot\boldsymbol{\sigma})$, $\rho_f=(1/2) (1+\mathbf{n}\cdot\boldsymbol{\sigma})$, $\Hat{A}=\mathbf{a}\cdot\boldsymbol{\sigma}$, with $\mathbf{a}$ a unit vector and $|\mathbf{m}|\le 1$, $|\mathbf{n}|\le 1$. 
The probability of a successful postselection is then
\begin{align}
\P_{post} &= \frac{1}{2}\left[ 1+\mathbf{m}\!\cdot\!\mathbf{a}\, \mathbf{n}\!\cdot\!\mathbf{a} 
+e^{-2\lambda^2/\kappa_k^2-2\lambda^2/k_H^4}\left(\mathbf{m}\!\cdot\!\mathbf{n}-\mathbf{m}\!\cdot\!\mathbf{a}\,\mathbf{n}\!\cdot\!\mathbf{a}\right)\right]
\end{align}
The joint probability is
\begin{align}
\nonumber
\P(k,\rho_f) &= \frac{1}{4}\biggl\{ (1+\mathbf{m}\!\cdot\!\mathbf{a})(1+\mathbf{n}\!\cdot\!\mathbf{a}) \P_0(k-\lambda)+ 
(1-\mathbf{m}\!\cdot\!\mathbf{a})(1-\mathbf{n}\!\cdot\!\mathbf{a}) \P_0(k+\lambda)
\nonumber
\\
&+2e^{-2\lambda^2/\kappa_k^2}\left[
\left(\mathbf{m}\!\cdot\!\mathbf{n}-\mathbf{m}\!\cdot\!\mathbf{a}\,\mathbf{n}\!\cdot\!\mathbf{a}\right)
\cos{\!\left(\frac{2\lambda k}{k_H^2}\right)}
-(\mathbf{m}\!\times\!\mathbf{n})\!\cdot\!\mathbf{a}\, \sin{\!\left(\frac{2\lambda k}{k_H^2}\right)}
\right]\P_0(k)\biggr\}.
\label{eq:probspin}
\end{align}

The characteristic function is
\begin{align}\nonumber
Z(\theta) &= \frac{Z_0(\theta)}{\P_{post}}\frac{1}{2}\biggl\{  
\left(1+\mathbf{m}\!\cdot\!\mathbf{a}\, \mathbf{n}\!\cdot\!\mathbf{a}\right)\cos{(\lambda\theta)} 
+i\left(\mathbf{m}\!\cdot\!\mathbf{a}+ \mathbf{n}\!\cdot\!\mathbf{a}\right)\sin{(\lambda\theta)}
\\
&+e^{-2\lambda^2(\kappa_k^{-2}+k_H^{-4})}\left[
\left(\mathbf{m}\cdot\mathbf{n}-\mathbf{m}\!\cdot\!\mathbf{a}\,\mathbf{n}\!\cdot\!\mathbf{a}\right)
\cosh{\!\left(\frac{2\lambda\theta}{k_H^2}\right)}
-i (\mathbf{m}\!\times\!\mathbf{n})\!\cdot\!\mathbf{a} \, 
\sinh{\!\left(\frac{2\lambda\theta}{k_H^2}\right)}
\right]\biggr\},
\end{align}
with $Z_0(\theta)=\exp{[-\theta^2/2]}$. 
\subsection{Coherent oscillations}
Equation \eqref{eq:probspin} shows that the off-diagonal terms of $\rho_0(k,k\rq{})$ provide oscillating terms, with a period 
$k_{osc}=\pi k_H^2/\lambda$. If $k_{osc}$ exceeds the scale $\Delta_k$ over which $\P_0(k)$ decays, the oscillations can not be discerned. If instead $k_{osc}\ll \Delta_k$, coherent oscillations are observed. In this case, we can not expand in $\lambda$
the exponential of the Hamiltonian phases $\exp{(i\Gamma_a)}$, but should retain the full terms. 
This can be implemented through the transformation $\rho_f\to U_k\rho_f U^\dagger_k=\rho_f(k)$, 
with the unitary operator 
\begin{equation}
U_k=\exp{\left\{i \lambda\omega_P\rq{}(k)\tau_0\Hat{A}-i\frac{\lambda^2}{2}\omega_P''(k)(\tau_0-\tau_1)\Hat{A}^2\right\}}.
\end{equation}
For a spin 1/2, $U_k$ is but a rotation around the direction of $\Hat{A}$. 
Accordingly, the weak value becomes a periodic function of $k$
\begin{equation}
A_w=\frac{\Tr\{\rho_f(k) \Hat{A}\rho_i\}}{\Tr\{\rho_f(k) \rho_i\}}\, , \  
B_w=\frac{\Tr\{\rho_f(k) \Hat{A}\rho_i\Hat{A}\}}{\Tr\{\rho_f(k) \rho_i\}}.
\end{equation}
Furthermore $x_t\to x$ and $x_{t|k}\to x$ in Eqs.~(10)-(14). 
Once these prescriptions are applied, the probability in Eq. (13) approximates excellently the exact expression.

\end{document}